\documentclass[useAMS,usenatbib]{mn2e}
\usepackage{graphicx}
\usepackage{amssymb,amsmath,multirow}
\usepackage{epstopdf,enumerate,url,color}
\usepackage{fixltx2e} 
\title[Torque on a planet from an evaporative wind]{Torque on an exoplanet from an anisotropic evaporative wind}
\author[Teyssandier et al.]{Jean Teyssandier$^{1}$, James E. Owen$^{2,3,7}$, Fred C. Adams$^{4,5}$, \& Alice C.  Quillen$^6$\\
$^1$ Department of Applied Mathematics and Theoretical Physics, University of Cambridge, Cambridge CB3 0WA, United Kingdom \\
$^2$ Canadian Institute for Theoretical Astrophysics, 60 St. George Street, Toronto, Ontario M5S3H8, Canada \\
$^3$ Institute for Advanced Study, Einstein Drive, Princeton, NJ 08540, USA\\
$^4$ Michigan Center for Theoretical Physics
Physics Department, University of Michigan, Ann Arbor, MI 48109, USA \\
$^5$ Astronomy Department, University of Michigan, Ann Arbor, MI 48109, USA     \\
$^6$ Department of Physics and Astronomy, University of Rochester, Rochester, NY 14627, USA\\
$^7$ Hubble Fellow    
}

\begin{document}
\maketitle

\begin{abstract}

Winds from short-period Earth and Neptune mass exoplanets, driven by high energy radiation from a young star, may evaporate a significant fraction of a planet's mass. If the  momentum flux from the evaporative wind is not aligned with the planet/star axis, then it can exert a torque on the planet's orbit. Using steady-state one-dimensional evaporative wind models we estimate this torque using a lag angle that depends on the product of the speed of the planet's upper atmosphere and a flow timescale for the wind to reach its sonic radius. We also estimate the momentum flux from time-dependent one-dimensional hydrodynamical simulations. We find that only in a very narrow regime in planet radius, mass and stellar radiation flux is a wind capable of exerting a significant torque on the planet's orbit. Similar to the Yarkovsky effect, the wind causes the planet to drift outward if atmospheric circulation is prograde (super-rotating) and in the opposite direction if the circulation is retrograde. A close-in super Earth mass planet that loses a large fraction of its mass in a wind could drift a few percent of its semi-major axis. While this change is small, it places constraints on the evolution of resonant pairs such as Kepler 36 b and c.

\end{abstract}


\section{Introduction}
\label{sec:intro}

Planets discovered by the Kepler Observatory \citep{batalha13} exhibit a large range
of densities ranging from less than 1 g~cm$^{-3}$
(e.g. Kepler 36c; \citealt{carter12} and Kepler 79d, \citealt{jontofhutter14}) 
to greater than 8 g~cm$^{-3}$ (e.g. Kepler 10b, \citealt{batalha11}).
For Neptune and Earth mass planets, the inferred gaseous envelope masses range from zero to tens of percent of the total planetary mass, with
this fraction being dependent on whether a rock/iron or ice core is adopted \citep{howe14}.
Either close-in planets are formed with a wide range of core to envelope mass ratios,  
or their compositions evolve after they have accreted gas (and after evaporation of the circumstellar disk).
Evaporative winds  driven by stellar UV and X-ray radiation  could account for
the loss of a significant fraction of a Neptune-sized planet's gaseous envelope mass 
\citep{lopez13,owen13}.  
Impacts or collisions between bodies that are more frequent at small semi-major axis
 can also affect the planetary composition (e.g., \citealt{quillen13}).
Among planet pairs in multiple-planet systems, a larger fraction ($\sim 60\%$) of
systems have the smaller radius planet residing closer to the star \citep{ciardi13}. Along with the closer in planet generally being denser \citep{wu13}.
As more mass can be evaporated closer to a star, evaporative winds could account for this trend \citep{owen13}.

In addition, evidence for mass loss through atmospheric evaporation have been observed for the hot Jupiters around HD 209458  \citep{vidalmadjar03} and  HD 189733 \citep{lecavelier10,bourrier13}, with loss rates estimated to be on the order of $10^{10}\ {\rm g~s}^{-1}$.

An evaporative wind from an exoplanet is often assumed to be isotropic.  However, 
an evaporative wind carries momentum and could exert a torque on the planet
affecting both its orbit and spin. Here we first summarize related physical phenomena.
\citet{veras13} showed that an anisotropic stellar wind can cause orbital inclination changes.
The Yarkovsky effect occurs when uneven re-emission of absorbed stellar radiation causes drifts in asteroid
semi-major axes (see \citealt{bottke06}).  
It has also been shown that sublimation from a comet can exert a reactive torque on the cometary nucleus \citep{gutierrez02,neishtadt03}.
\cite{boue12} proposed that an anisotropy in an evaporative wind from a close-in Jupiter- or Neptune-mass planet 
can cause the planet's orbit to migrate 
(see also \citealt{iorio12}). \cite{boue12} suggested that migration distances inferred from distributions of planetary mass
and orbits could be used to constrain the launch angle of the evaporative wind.  In this paper we attempt to estimate
the launch angle  of the wind and the magnitude of the resulting reactive torque on the planet based on hydrodynamical models. 
As evaporation is primarily significant for low core mass planets experiencing high UV and X-ray fluxes \citep{lopez13,owen13}
from the star, we focus here on winds driven from warm Neptunes and super-Earth like planets within 0.2 AU of a star.

\begin{figure*}
\begin{center}
\includegraphics[width=0.8\textwidth]{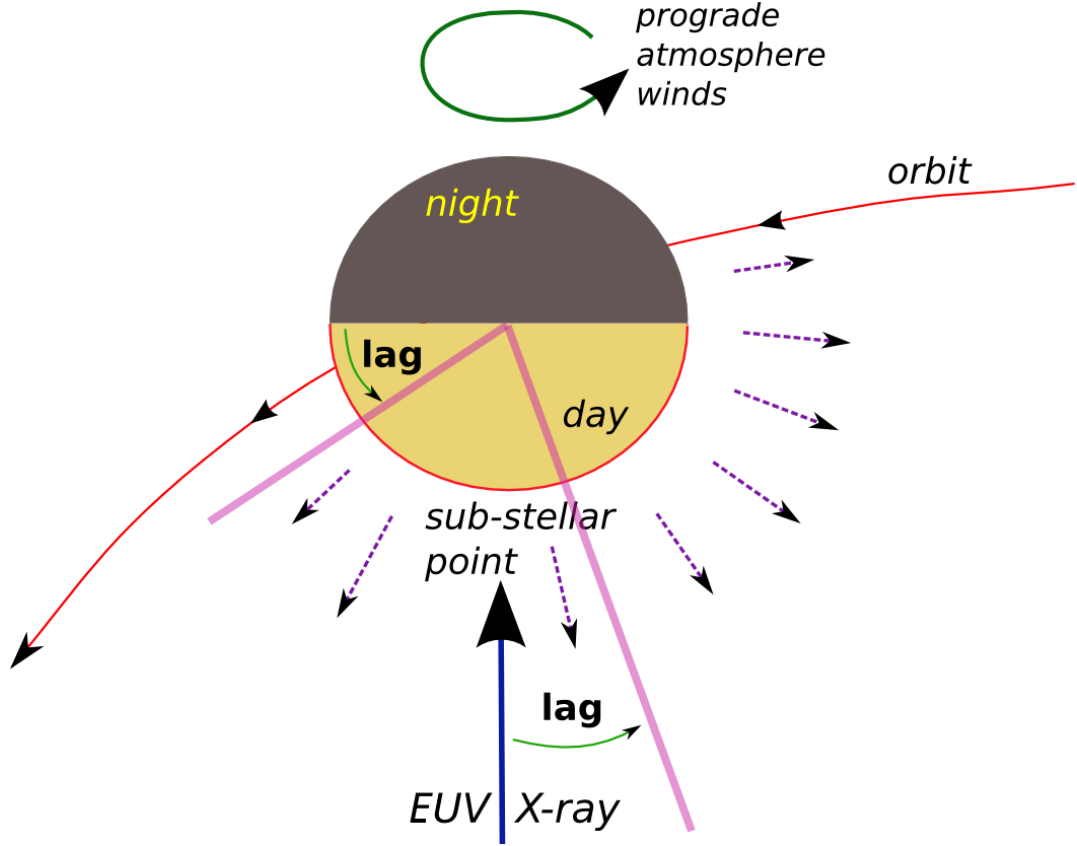}
\caption{
If the wind launch region is super-rotating with respect to the planetary surface, then the wind is launched after the upper atmosphere sees dawn. A comparison between a timescale to launch the wind and with the time required for the upper atmosphere to rotate gives an angular lag in the peak of the day-driven wind. Because of the lag, the momentum flux of the wind is offset compared to the star/planet axis and this gives a torque on the planet.  If the upper atmosphere is super-rotating in the prograde direction with respect to the planetary surface, the torque increases the angular momentum of the planet and the planet's orbit moves outward.
\label{fig:wind_pic}
}
\end{center}
\end{figure*}

As the cooling timescale in an ionosphere is expected to be shorter than a day, an evaporative wind is mainly expected to be driven from
the day side of a planet.  If the planet is not tidally locked
or if the atmosphere is circulating then the upper atmosphere would experience day/night variations
in stellar illumination.  A delay between dawn and the time it takes for the wind to develop
could cause an asymmetry in the wind (see Figure \ref{fig:wind_pic}).
Here we aim to estimate the torque on a planet due to this possible asymmetry.   

Planets within 0.2 AU are expected to be tidally locked and at low obliquity (e.g., \citealt{lubow97,cunha14}). 
Spin-down should have taken place in less than $\sim 10^4$ years (see Table 1 by \citealt{cunha14}), 
and well before the time when most of the evaporation takes place, 
at an age $\sim 10^7 -  10^8$ years \citep{owen12}. 
However, some planets could now be in a 
 spin synchronous state (with rotation rate related to the angular rotation rate at pericenter,  \citealt{hut80}) 
 or spin-orbit resonance \citep{rodriguez12} like Mercury.  
Even though the tidal spin-down timescale is short, the timescale for tidal circularization of the orbit  is relatively 
long (e.g., \citealt{ford08,lee13}) and planets could have been at moderate (or their current) eccentricities
when atmospheric evaporation took place.   About a quarter of the planets in multiple-planet Kepler systems with transit-timing variations
have high eccentricities, in the range e=0.1-0.4   \citep{wu13}. 
In exoplanetary systems containing at least two planets,  transit timing variations are consistent 
 with an eccentricity dispersion of $0.018^{+0.005}_{-0.004} $ when fit with a Rayleigh distribution \citep{hadden14}.

Even if a planet is tidally locked, the upper atmosphere can be rotating
with advective or zonal flow speeds of order $v_{zonal} \sim 1$~km/s
\citep{showman11,menou12,rauscher14}.
If the advective flow speed is of order the sound speed (e.g., \citealt{cowan11b})  then we expect a wind
speed $\propto T_{eq}^{1/2}$, where $T_{eq}$ is the equilibrium temperature set by
radiation balance from the stellar light.   As this is weakly dependent on the planet's semi-major axis ($T_{eq} \propto a_p^{-1/2}$
with $a_p$ the semi-major axis of the planet), for simplicity
we can  neglect the possible dependence of advective flow speed on stellar flux. We also neglect the dependence on planet properties 
such as atmospheric opacity  and surface gravity, as atmospheric models all predict similar circulation speeds.
 Evaporative winds are driven from a tenuous region of a planetary atmosphere 
 where ionization and heating occur, and this layer is higher than typically modelled to predict atmospheric circulation.
 Nevertheless, $\sim$1 km/s size-scale zonal flows are  predicted in exoplanet ionospheres (see Figure 3 by \citealt{koskinen10}).
 
For completeness, we note that, in addition to the effects outlined above, magnetic fields on the planetary surface will lead to anisotropy. More specifically, with the expected mass loss rate from planets and surface field strengths of $\sim$ 1 Gauss, the outflows could be magnetically controlled \citep[see also \citet{adams11}]{owen14}. This issue must be addressed in future work.
 
Both zonal circulation or a spin-orbit resonance would cause a planetary ionosphere to
experience day/night variations. In section \ref{sec:rough} we estimate a torque due to an evaporative wind by considering
the relationship between atmospheric motion and the timescale to launch a wind at dawn, as this is expected to dominate the rotation of gas in the upper atmosphere, unless the planet is in an unusual spin-orbit resonance.
In section \ref{sec:theta} we compute a lag angle for the wind momentum loss rate using one-dimensional 
hydrodynamical steady-state
evaporative wind models  \citep{owen12}.  
In section \ref{sec:hydro} we compute this angle by integrating one dimensional but
time-dependent hydrodynamic models.  Lastly in \ref{sec:kep36}  we investigate scenarios for the Kepler 36 
planetary system taking into account an evaporative wind induced torque.  
Variations in the planet's orbital elements caused by a torque from an anisotropic evaporative wind, integrated over the orbit, are computed in the appendix.

\section{Torque estimate and Lag angle}
\label{sec:rough}

In this section we roughly estimate a lag angle and the torque exerted by the wind onto the planet. We first derive a launch timescale for the wind, and compare it to the circulation timescale in the upper atmosphere.

\subsection{Relevant timescales}

Stellar X-ray and EUV radiation heats gas up to temperatures $T \sim 10^4$ K  \citep{spitzer78,shu92,ercolano08} 
corresponding to a sound speed of  $c_{EUV} \sim 10$ km/s.  Above this temperature radiative cooling is so efficient
that higher temperatures are rarely reached.
As $c_{EUV}$ is similar to a planet's escape velocity, a wind can be launched from the heated gas.
An isothermal or Parker-type wind model \citep{parker58,parker65} gives an approximate hydrodynamical
description for evaporative winds from close-in planets, i.e., planets for which evaporative winds can be important \citep[out to about 0.2 AU, see, e.g.,][]{owen12}. 
Such a wind cannot be instantaneously launched at dawn.  
We can estimate a launch timescale for the  wind from the time it takes a particle in the wind to travel 
from the planet's surface, at a planetary radius $R_p$, to the transonic radius, $R_s$ (where the flow becomes supersonic), and is no longer in causal contact with the planet. 
\begin{equation}
t_{flow} \sim \int_{R_p}^{R_s} \frac{dr}{u} \sim  \frac{R_s}{c_{EUV}}
\end{equation}
where $u$ is the wind velocity as a function of radius.
The one-dimensional hydrodynamic models that include ionization and heating by X-rays, \citet{owen12}  have  
  $R_s \sim 2-3 R_p$ (see their Figure 4) and $c_{EUV}\sim5$~km~s$^{-1}$, where $R_p$ is the planet's radius.  Using these larger transonic radii we find:
 \begin{equation}
  t_{flow} \sim  \frac{R_s}{c_{EUV} }\sim  10^3~s \left( \frac{R_p}{R_\oplus} \right)  
  \left( \frac{ R_s/R_p }{2} \right)
  \left(\frac{c_{EUV} }{5 ~{\rm km~s}^{-1}} \right)^{-1}.
    \label{eqn:t_flow}
 \end{equation}

The motion of the upper atmosphere wind, described by a zonal wind speed $v_{zonal}$, and 
 the timescale to launch the wind together imply that the wind is not launched at dawn but slightly afterwards.
It would take a similar time for the evaporative wind to die down after dusk.
 The delay implies that the average wind momentum is not pointing exactly toward the star along the sub-stellar
 point but is shifted by an angle $\theta_{lag}$ from this direction (see Figure \ref{fig:wind_pic}).
We estimate this angle to be
\begin{equation}
\theta_{lag} \sim \frac{ t_{flow} }{t_{zonal} }, 
\end{equation}
where
\begin{equation}
t_{zonal} \sim \frac{R_p}{v_{zonal}}  = 637 {\rm s}
 \left( \frac{v_{zonal}}{1 {\rm km~s}^{-1}} \right)^{-1} \left( \frac{R_p }{R_\oplus} \right) \label{eqn:t_zonal}
\end{equation}
is the inverse of the atmosphere's angular rotation rate.
A comparison between equation (\ref{eqn:t_zonal}) and equation (\ref{eqn:t_flow}) suggests
that the angle $\theta_{lag}$ could be of order 1 and the wind could exert a significant torque on the planet.

If  $t_{flow} \gg R_p/v_{zonal} $ then we could consider the wind launched evenly from both day and night sides. 
If $t_{flow} \ll R_p/v_{zonal}$ then the momentum flux from the wind (averaged over all regions on the planet surface) is in the radial direction toward the star, and there is no torque exerted by the wind onto the planet.
We have used a flow timescale to estimate the time it takes to launch the wind that is based on a steady state flow model.  
An alternative thermal timescale, $t_{kh}$, that
can be used to estimate a wind launch timescale 
is the time it takes to replenish the kinetic energy in the flow within the sonic radius from the absorbed EUV and X-ray
energy flux.   Depending upon the ratio of the transonic to planet radius, and as we discuss below, 
a thermal timescale may more accurately describe
the time it takes to launch the wind. We detail this possibility in section \ref{sec:theta}.
We have neglected the wind speed's tangential velocity component in our estimate for $\theta_{lag}$ as $v_{zonal} $ 
is significantly smaller
than the escape velocity from the planet surface.

\subsection{Orbital and spin torques exerted by the wind}

The orbital torque (from the wind) can be roughly estimated from the lag angle $\theta_{lag}$ as
\begin{equation}
\tau_w \sim \dot M \theta_{lag} v_w a_p/2 \label{eqn:tau_w}
\end{equation}
where $\dot M$ is the wind mass outflow rate and $v_w \sim c_{EUV}$ is a mean wind velocity near the transonic region.
The rough factor of 2 comes from integrating a radial wind over a hemisphere with an angular offset.

The sign of the lag is set by the direction of the planet's spin or atmospheric flow.  
If the planet's spin or the atmospheric circulation are prograde (super-rotating or eastward)
then the torque is  positive and we expect the planet to move outwards.
The sign is in direct analogy to the Yarkovsky effect (see \citealt{bottke06}), 
with outward migration in semi-major axis for prograde asteroid body rotation
and vice versa for retrograde body rotation.
Close-in exoplanet 
atmospheric circulation models usually predict prograde (super-rotating or eastward)
equatorial upper atmospheric circulation \citep{showman11,menou12,rauscher14},
with the exception of the retrograde (westward) slowly rotating model for HD 209458b by \citet{rauscher14}. Although, we caution that the atmospheric rotation of low-mass close in exoplanets has not been studied and note in passing that the circulation direction in Neptune and Uranus is retrograde.

Here we also note that the evaporative wind could affect the spin evolution of the planet. The torque on the planet itself we estimate as
\begin{equation}
\tau_s \sim \dot M (v_{zonal} - R_p \Omega_p ) R_p
\end{equation}
where $v_{zonal}$ is the atmosphere circulation velocity and we assume that the wind is launched at a radius
that is not significantly different than $R_p$.
Here $\Omega_p$ is the spin of the planet.  Assuming that $v_{zonal} > R_p \Omega_p$
and after time $\Delta t$, one gets a change in planet spin of 
\begin{equation}
\Delta \Omega_p \sim \frac{ \dot M \Delta t}{M_p} \frac{v_{zonal}}{R_p} \frac{1}{\alpha_p}
\end{equation}
where the planet's moment of inertia is $I_p = \alpha_p M_p R_p^2$. For short-period, tidally locked planets, this effect could act to desynchronise the planet's spin, provided synchronisation occurred before the evaporation became important.

\subsection{Orbital evolution}

A strong evaporative wind is driven when the host star has a strong 
EUV and X-ray flux or during an age less than $\Delta t \sim 10^8$ years
\citep{owen12}.  During this time, a planet of mass $M_p$ can experience a loss of mass $\Delta M_p = \dot M \Delta t$
and a change in semi-major axis $\Delta a_p$.
The change in the planet's orbital angular momentum is
\begin{equation}
\Delta L \approx \tau_w \Delta t \sim { M_p v_c \Delta a_p}{/2},
\end{equation}
where $M_p$ is the planet mass and $v_c$ the planet's orbital velocity (assuming a nearly circular orbit), which scales as
\begin{equation}
v_c = 94 {\rm km~s}^{-1} \left(\frac{M_*}{M_\odot} \right)^\frac{1}{2}
 \left(\frac{a_p}{0.1~{\rm AU}} \right)^{-\frac{1}{2}} .
\end{equation}
Using equation (\ref{eqn:tau_w}) for the torque, we estimate 
\begin{equation}
\frac{ \Delta a_p }{a_p} \sim \frac{\dot M \Delta t}{M_p} \frac{v_w}{v_c}  \theta_{lag}
\sim \frac{\Delta M_p}{M_p}\frac{v_w}{v_c}  \theta_{lag}\label{eqn:da}
\end{equation}
The factor $v_w/v_c$ limits the size-scale
of the possible migration distance in semi-major axis.

Many of the Kepler planet host stars have masses lower than $1M_\odot$ \citep{batalha13} and for these lower mass
stars the ratio $v_w/v_c$ would be higher.
Previous studies \citep{owen13,lopez13} have found that close-in low core mass planets can loose a significant
fraction of their mass.   Taking $\Delta M_p/M_p \sim 1$ and $\theta_{lag} \sim 1$, 
Equation (\ref{eqn:da}) gives
\begin{equation}
\frac{ \Delta a_p }{a_p} \sim 0.1
\frac{\Delta M_p}{M_p}\theta_{lag} \left( \frac{v_w}{10 ~{\rm km~s}^{-1} }\right) 
\left(\frac{M_*}{M_\odot} \right)^\frac{1}{2}
\left(\frac{a_p}{0.1~{\rm AU}} \right)^{-\frac{1}{2}}.
\end{equation}
A planet that loses a significant fraction of its mass could move a significant fraction (say 10\%) of its semi-major axis.
 10\% exceeds the distance to resonance exhibited by resonant pairs \cite{fabrycky14} and is dynamically significant,
 particularly for close resonant pairs such as Kepler 36b,c \citep{carter12}.
 
In appendix \ref{ap:orbital}, we derive more general equations for the secular evolution of semi-major axis and eccentricity for a planet whose orbit is perturbed by an evaporative wind.

\section{Lag Angle estimated from steady-state models}
\label{sec:theta}

In the previous section we estimated the torque on  a planet due to evaporative wind loss
using a crudely approximated launch timescale for the wind.
In this section we use the steady-state hydrodynamic models, including radiative cooling and ionization, by \citet{owen12} to better estimate the lag angle $\theta_{lag}$ that sets the torque from the wind.

\begin{figure*}
	\begin{center}
		\includegraphics[scale=0.5]{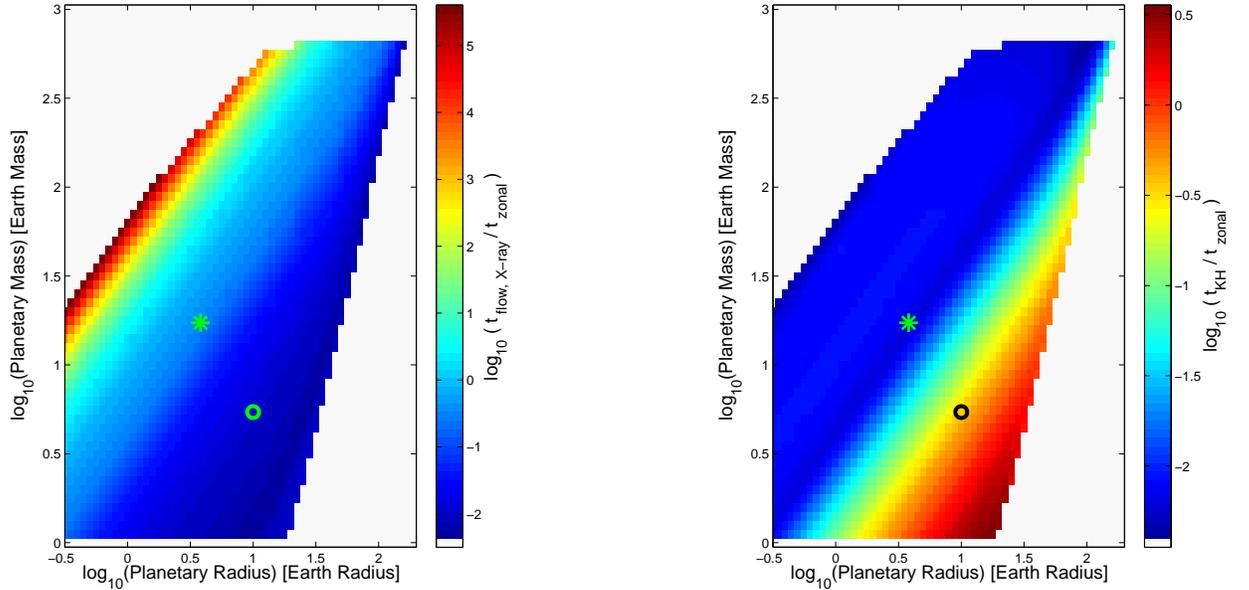}
		\caption{\textit{Left:} $t_{flow}/t_{zonal}$ as given by eq. (\ref{eq:tflow}) for an X-ray + EUV stellar luminosity of $10^{30}$ erg/s and a semi-major axis of 0.1 AU, using the results of numerical simulations presented in \citet{owen14}. \textit{Right:} same for $t_{kh}/t_{zonal}$ as given by equation (\ref{eq:tkh}). In both plots, the star represents Neptune, while the circle represents a pre-evaporation Kepler 36b.}
		\label{fig:tflowtkh}
	\end{center}
\end{figure*}

As discussed above we take the flow timescale to be the time to establish an evaporative flow. It is the time the flow takes to reach the sonic point and is thus found by:

\begin{equation}
\label{eq:tflow}
t_{flow}=\int_{R_p}^{R_s}\frac{\text{d} r }{u}
\end{equation}
where $u$ is the velocity field of the flow. The flow solutions from \citet{owen12} allow us to directly perform this calculations without making further assumptions. A comparison between this flow timescale and the zonal wind timescale using a zonal wind velocity of 1~km~s$^{-1}$ (Equation~[\ref{eqn:t_zonal}]) is shown in the left-hand panel of Figure~\ref{fig:tflowtkh}. In line with our discussion in Section~2 the flow timescale is of order the zonal flow timescale for Neptune like planets.  

One further consideration is that the flow receives enough energy to launch the wind, we call this timescale the ``Kelvin-Helmhotlz timescale'' of the flow (following Owen \& Adams, 2014). We can also estimate the time it takes to heat the wind from a  ratio of kinetic energy (integrated out
to the sonic radius,  $\sim \rho_s c_{s}^2 4\pi R_s^3/3$) 
and the energy absorbed per unit time in UV and X-ray photons, $\sim \pi R_p^2 F_{UV}$
where $\rho_s$ is the density in the wind at $R_s$ and $F_{UV}$ is the flux in X-ray and EUV photons from the star. 
In analogy to the Kelvin-Helmholtz timescale we write
\begin{equation}
\label{eq:tkh}
t_{kh} = \frac{ \int_{R_p}^{R_s} 4\pi r^2 \text{d}r \rho u^2} {\pi R_p^2 F_{UV}},
\end{equation}
where the numerator represents the integral of the energy density in the flow over the volume within the sonic radius. We show the calculated ``Kelvin-Helmhotz timescales'' compared to the zonal flow timescale in the right panel of Figure~\ref{fig:tflowtkh}. We find that the flow timescale dominates the launch timescale when the potential is deep (as the suppression of the launch velocity is exponentially sensitive to planet mass) and the flow is not close to ``energy-limited'' (see discussion in Owen \& Jackson 2012). However, when the potential is shallow, the sonic point is very close to the planet's surface and the flow timescale is essentially zero. In this case the timescale to launch the wind is controlled by the energy considerations. 

\begin{figure}
	\begin{center}
		\includegraphics[scale=0.5]{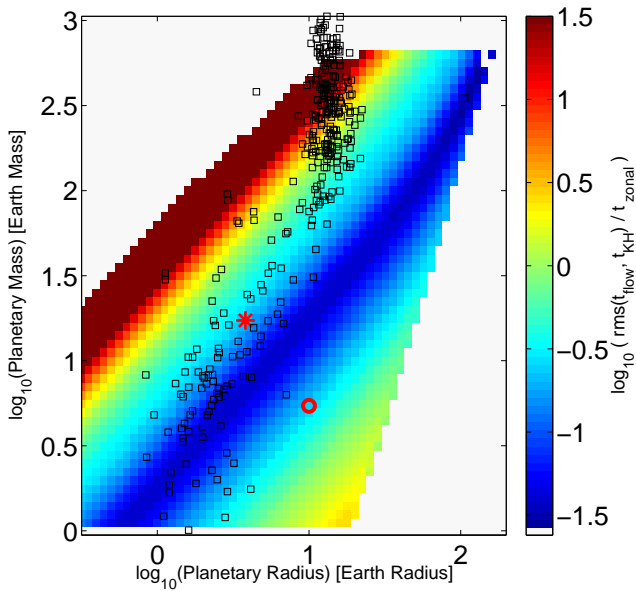}
		\caption{Root mean square of $t_{flow}/t_{zonal}$ as given by eq. (\ref{eq:tflow}) for an X-ray + EUV stellar luminosity of $10^{30}$ erg/s and a semi-major axis of 0.1 AU, using the results of numerical simulations presented in \citet{owen14}, and $t_{kh}/t_{zonal}$ as given by equation \ref{eq:tkh}. As a guide, we also plot a sample of exoplanets with orbital periods less than 100 days. In addition, the star represents Neptune, while the circle represents a pre-evaporation Kepler 36b.}
		\label{fig:tflowrms}
	\end{center}
\end{figure}

In reality both timescales matter. Therefore we define a ``launch timescale'' as the root-mean square of the tow timescales, and we compare it to the zonal flow timescale. This comparison is shown in Figure~\ref{fig:tflowrms}. It shows that there is a reasonably wide band of planet parameters where the launch timescale compared to the zonal timescale is order unity. We compare it to the existing sample of exoplanets. We use the database from the Open Exoplanet Catalogue\footnote{http://www.openexoplanetcatalogue.com/} \citep{rein12}, for all planets with orbital period lower than 100 days. A small fraction of the super-Earth to Netpune-like planets are in the range where the timescale ratio (and hence our estimated lag angle) is of order unity.

\section{Momentum loss rate estimated from one-dimensional time dependent hydrodynamical models}
\label{sec:hydro}

To improve on the estimates of section \ref{sec:theta}
we carry out some one dimensional but time-dependent hydrodynamic models.  From these we directly estimate the lag
angle and the vector direction of the momentum carried by the evaporative wind.

\subsection{Method}
We conduct a set of simulations using \textsc{zeus} \citep{stone92}. The code we us includes modifications for UV ionization described in detail by Owen \& Adams (2014) as well as time-dependant heating and cooling described in detail by \citet{owen2015}. The numerical method takes into account ionization by a range of EUV photons, and as such includes photon-hardening using the method of \citet{mellema06}, assuming a pure Hydrogen atmosphere, where ray-tracing is performed in a causal manner starting at the edge of the grid and working towards the planet. Heating is dominated by photo-ionization and cooling from Lyman-$\alpha$ and recombinations. 3D simulations of this problem are beyond the scope of this work and therefore we chose to setup a simple illustrative problem, that combined with the insight from our discussion above, allows us to make inferences about the role of evaporation in driving a planetary torque.

Our setup is as follows: we consider a small patch of the planet rotating with azimuthal velocity $v_{\phi}$ such as $v_{\rm zonal}=v_{\phi}\cos\theta$
where $\theta$ is the latitude of the patch. For each patch we solve the 1+1D (including the evolution of $v_{\rm \phi}$) in spherical co-ordinates. We note that this implicitly assumes that the flow has spherical divergence along the streamline and we neglect any pressure forces perpendicular to the radial direction. This is obviously a simplification; however for a thermally driven outflow the pressure forces in the radial direction must dominate. During the calculation we then modulate the input UV flux as a function of time to mimic the patch rotating from the day to night side and wait to a quasi-repetitious flow is established, after several rotation periods of the patch. Once we have simulated different patches at different latitudes we can then ``stitch'' the flow patterns together to gain an estimate of what the 3D flow structure is like. In particular, we can then explicitly estimate what the net torque on the planet is. As we have stitched together several patches one loses the constrain the momentum flux will be conserved in a global 3D sense; and as such we must be cautious in interpreting our results. 

The grid is radially divided in 128 cells, and the time-step is such that it takes about 100 iterations for a patch to go around the planet. In total there are 61 patches going from south pole to north pole. The outer boundary of the radial grid is such that all the flow becomes supersonic inside the grid.

\subsection{Results}
First, we study a Neptune-like planet, with a radius of $2.4\times 10^9$ cm and a mass of $1.03 \times 10^{29}$ g. It is irradiated with a flux of $2\times 10^{6}\ \text{erg s}^{-1} \ \text{cm}^{-2}$. This corresponds to a planet at 0.013 AU from a star with a X-Ray and UV luminosity of $10^{30}\ \text{erg s}^{-1}$, i.e. a young solar-type star. 
In what follows, we also note $\rho$, $P$ and $u$ the density, pressure, and velocity fields of the flow.

\begin{figure*}
    \begin{center}
    \includegraphics[width=0.8\textwidth]{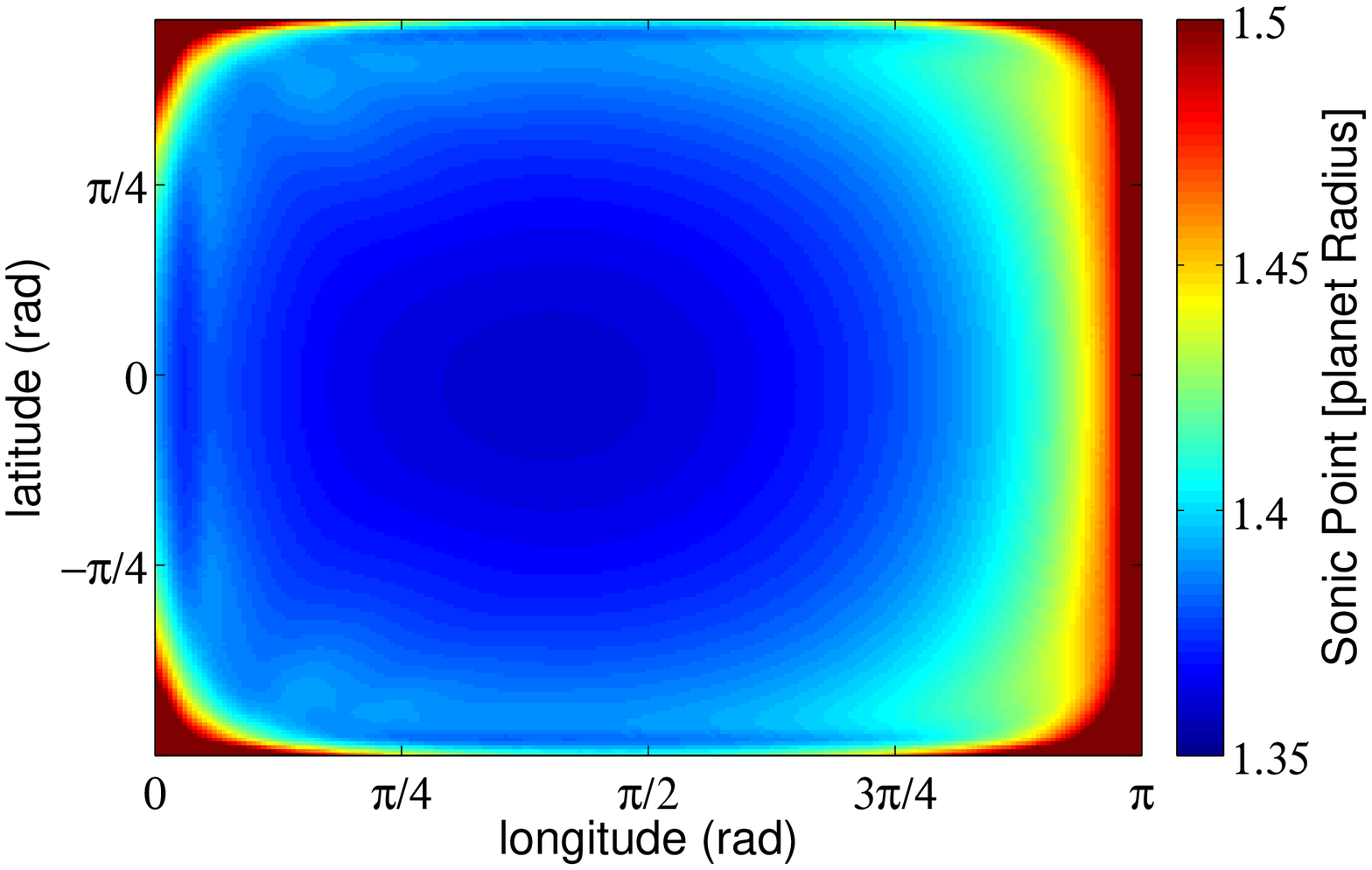}
    \caption{2D map of sonic radius divided by planetary radius for a Netpune-like planet, as a function of longitudes (day-side only, 0 is dawn) and latitudes (from south pole to north pole, 0 is at equator). The minimum sonic radius is offset from the substellar point by an angle of approximatively $16^{\circ}$ westward.}
    \label{fig:sonicmapnept}
    \end{center}
\end{figure*}

In figure \ref{fig:sonicmapnept} we show the sonic point for different latitudes and longitudes, on the day-side only. As expected, the sonic point is close to the planetary surface around the substellar point. However, the minimum of the curve appears to  be offset from the substellar point by a few degrees. We estimate that, at the equator, the minimum sonic radius is offset from the sub-stellar point by $16^{\circ}$ westward.

From this set of simulation, we can also extract the total mass loss, from our combined stitched together 3D flow solution, integrated over the longitudes $\phi$ and latitudes $\theta$, at an altitude $r$ (e.g., the sonic radius). It reads:
\begin{equation}
\dot{M} = \int_{\theta=-\pi/2}^{\pi/2}\int_{\phi=0}^{2\pi}\sin\theta \text{d}\theta \text{d}\phi \rho(r,\theta,\phi) u(r,\theta, \phi)r^2.
\end{equation}
In this case we find it to be $1.3\times 10^{12}\ \text{g s}^{-1}$, similar to the 1D spherically symmetric mass-loss rate of \citet{owen12}. 

Another relevant quantity is the the momentum loss rate of the wind, $\dot {\bf P}$, whose components $i=x,\ y, \ z$ are given by:
\begin{equation}
\dot{P}_i=-\int_{\partial V}\rho u_iu_j{\rm d}A^j - \int_{\partial V}P{\rm d}A_i -\int_V {\rm d}V\rho\partial_i\psi_g,
\label{eq:momflux}
\end{equation}
where $\psi_g$ is the gravitational field of the planet. $\dot {\bf P}$ is a vector, which we give in the following reference frame: the origin is at the substellar point, the $x$ direction lies along the line that goes from the substellar point the the star, the $y$ axis is perpendicular to the $x$ axis along the equator in the east direction, and the $z$ is perpendicular to the equator in the north direction. The problem is symmetric with regard to the planet's equator, so that the $z$ contribution cancels out once integrated over the whole planetary surface.

We estimate the lag angle of the wind $\theta_{lag}$ through the relation 
\begin{equation}
\tan\theta_{lag} = \dot{P}_y / \dot{P}_x.
\end{equation}
The integration is made in spherical coordinates, over our 3D stitched together flow solution, at a constant distance from the planet surface.

\begin{figure}
    \begin{center}
    \includegraphics[width=\columnwidth]{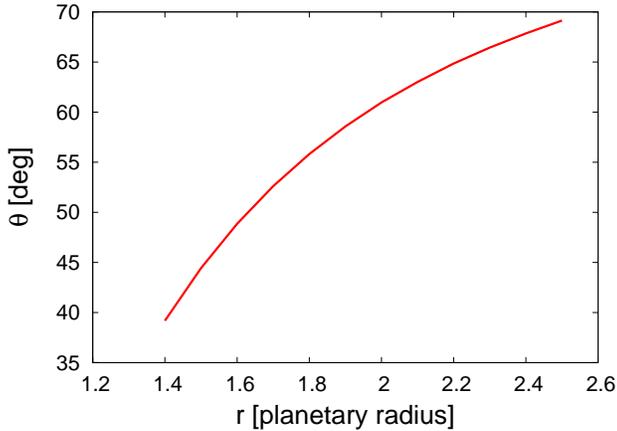}
    \caption{Lag angle $\theta$ as a function of the radial distance $r$ from the planetary surface, for a Neptune-like planet.}
    \label{fig:rtheta}
    \end{center}
\end{figure}

In the present simulations we find that the value of $\theta_{lag}$ depends on the distance at which the integration is conducted, as shown on figure $\ref{fig:rtheta}$. This is a consequence of the fact that our stitched flow solution does not conserve momentum explicitly in the angular directions as discussed above. In any case, we find that the mass loss is strongly anisotropic, with a lag angle of several tens of degrees.

Our one dimensional time-dependent model has therefore allowed us to show that a short-period, strongly irradiated, Neptune-like planet, looses mass at rate similar to other studies. In addition the momentum flux indicates that mass loss is anisotropic, and occurs at an angle that is offset from the substellar point by a few tenths of radians. According to our estimate in equation (\ref{eqn:da}), this would shift the semi-major axis of the planet by about 0.3\% after 100 Myr. In appendix \ref{ap:orbital} we also give a more general formula for the rate of change in semi-major axis due to the force exerted by the wind, which we can extract directly from the momentum flux. Using eq. (\ref{eqn:tau_a}) we find that over 100~Myr, the hot Neptune that we considered above would have migrated outward by about 1\% from the location at which we estimated the momentum flux (0.013 AU).

\section{Application to Kepler 36}
\label{sec:kep36}
Kepler 36b has a density $\rho_b = 7.46$ g~cm$^{-3}$
whereas Kepler 36c has $\rho_c = 0.89$g~cm$^{-3}$.
Kepler 36b is the lighter denser planet, suggesting that it could have lost much of its mass in a wind. The Kepler 36 exoplanet system is thought to have obtained such a dichotomy in densities due to evaporation \citep{owen13,lopez13}.
\subsection{Torque due to hydrodynamical escape}
If the torque on the planet was positive during evaporative wind loss then the planet would have migrated outward, pushing it closer to Kepler 36c. This would have pushed the system deeper into the 7:6 resonance rather than pulling it slightly out of resonance.

Currently, the system is just slightly outside of resonance with the separation between the two bodies larger than that of exact resonance \citep{carter12}. The two planets in this system could not have migrated much further apart as that would have made the system unstable \citep{deck12}.   

\citet{lopez13} have found that the peculiar architecture of the Kepler 36 system can be reproduced by evaporation of a pair of planets with cores of 4.45 $M_{\oplus}$ (inner planet) 7.34 $M_{\oplus}$  and (outer planet). Both cores are surrounded by a gaseous atmosphere, which initially contains  about 23\% of the total mass of the planet in the form of hydrogen and helium. Because of the close proximity of the pair of planets and the star, the atmospheres of the formers undergo hydrodynamical escape, mainly during the first $10^{8}$ yr, when the stellar flux is stronger. Since the core of the inner planet is less massive, it is unable to keep its atmosphere and only the dense rocky core remains, which we observe today. The outer planet manages to retain an extended gaseous envelope, resulting in an observed low mean density. The two planets of the Kepler 36 system are also very close to a 7:6 mean motion resonance. In this paper we study the possibility for the inner planet to have been slightly shifted out of the resonance under the action of the torque exerted by its own evaporating atmosphere.

We conduct hydrodynamical simulations similar to the ones presented above, for a planet consisting of a rocky core of 4.45 $M_{\oplus}$ and an atmosphere whose mass is 22\% of the core mass, and extends to 10 $R_\oplus$, in agreement to the model by \citet{lopez13}. It receives a flux of $2.2\times 10^{4}\ \text{erg s}^{-1} \ \text{cm}^{-2}$ at 0.12 AU. We the perform a series of 1+1D simulations to map out the flow structure using the method described in Section \ref{sec:hydro}, with the same resolution. 

\begin{figure*}
    \begin{center}
    \includegraphics[width=\textwidth]{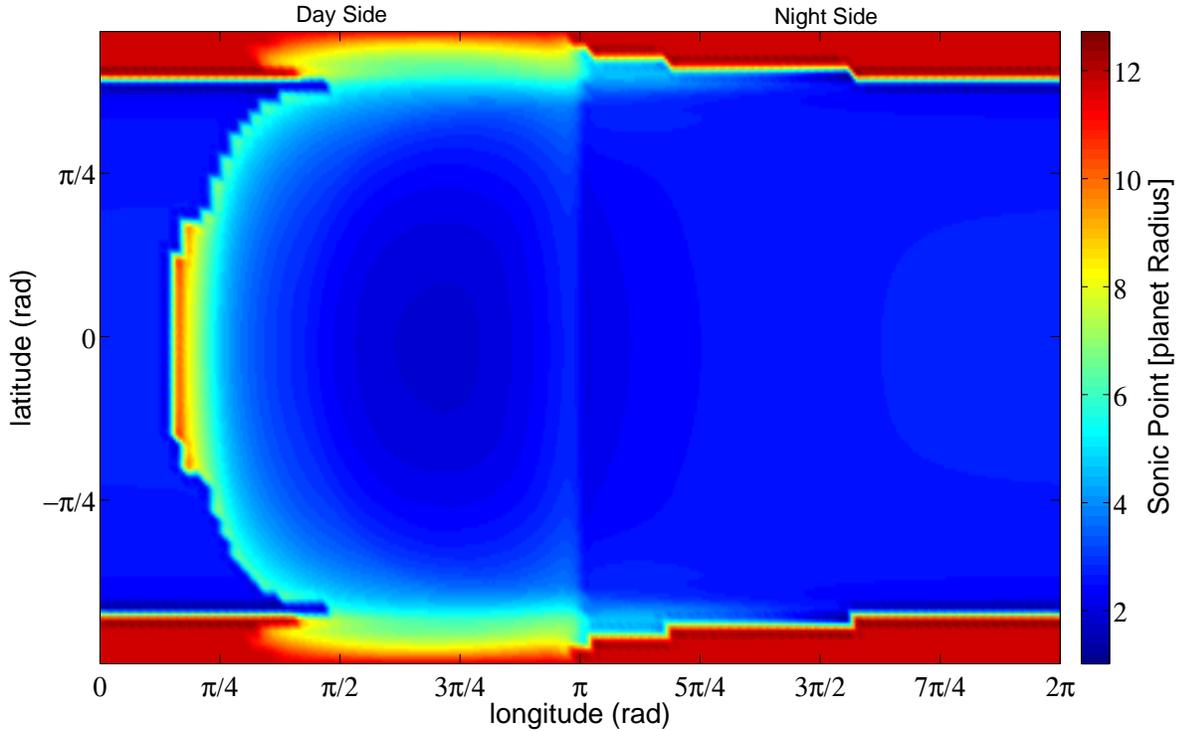}
    \caption{2D map of sonic radius divided by planetary radius for a pre-evaporation Kepler 36b planet, as a function of longitudes (day-side and night-side, 0 is dawn, $\pi$ is dusk) and latitudes (from south pole to north pole, 0 is at equator). }
    \label{fig:sonicmapk36}
    \end{center}
\end{figure*}

The sonic radius as a function of longitudes, at the equator, is presented in figure \ref{fig:sonicmapk36}, and includes both day and night sides. The minimum sonic point is largely offset from the substellar point, in the East direction. On the night side, the atmosphere does not entirely cool down and a wind still exists because of the temperature at the base of the atmosphere, which is set to 1000 K in our simulations. 

\begin{figure}
    \begin{center}
    \includegraphics[width=\columnwidth]{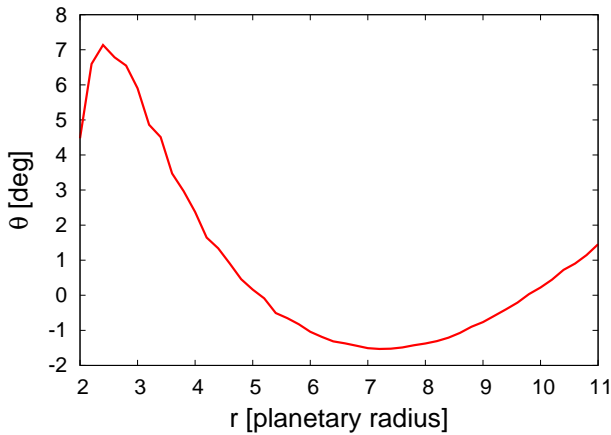}
    \caption{Lag angle $\theta$ as a function of the radial distance $r$ from the planetary surface, for a pre-evaporation Kepler 36b planet. The wind is rapidly launched in the day-side, and its density decreases at large radii. On the night side, there exists a quasi steady-state and isothermal wind, with a significant density at larger radii. This could explain the small, and sometimes negative, lag angle.}
    \label{fig:rthetak36b}
    \end{center}
\end{figure}

On figure \ref{fig:rthetak36b} we observe that the lag angle is minimum (and negative) at  intermediate radii ($r \sim 7~ r_{planet}$). In our simulations we remark that on the days-side, the wind is rapidly launched and decreases in density at large radii and never reaches a quasi steady state. On the night-side, there still exists a transonic point at which the wind is launched (see figure \ref{fig:sonicmapk36}). This wind is in quasi steady-state and isothermal, and has a larger density at large radii than the day-side wind. This can possibly explain the lag angle profile of figure \ref{fig:rthetak36b}. This is an example of a case where multi-dimensional simulation are required in order to fully understand the problem.

We find the mass loss to be $3.7\times 10^{12}\ \text{g s}^{-1}$. On figure \ref{fig:rthetak36b} we show that the lag angle strongly depends on the position at which it is evaluated, but does not exceed $8^{\circ}$. This lag angle is smaller than the one we found for a Neptune-like planet, which confirms the simple estimate we made in section \ref{sec:theta} (see left panel of figure \ref{fig:tflowtkh}).  Figure \ref{fig:sonicmapnept} clearly indicates that for the Neptune-like planet, the sonic point was located in a small range from 1.35 to 1.5 planet radius. In the case of Kepler 36, the sonic point spreads a much wider range of values. If we ignore the contribution from the polar regions, most of the region with $r/r_p>8$ is transonic. In this region, the lag angle increases with $r$, just as in the Neptune-like case.  From equation (\ref{eqn:da}) we estimate that the resulting drift in semi-major axis would be, at best, of 0.4\% after 100 Myr. This is fairly consistent with equation (\ref{eqn:tau_a}), which predicts an inward drift of 0.2\% after 100~Myr when the torque is maximum.

We have also conducted simulations for a pre-evaporation Kepler 36c planet, using again the model by \citet{lopez13}. We find that the lag angle increases from $15^{\circ}$ to $45^{\circ}$ degrees, as we increase the radial location at which the momentum flux is evaluated. The mass loss rate is $1.3\times 10^{12}\ \text{g s}^{-1}$. Using equation (\ref{eqn:da}) we estimate that the resulting drift in semi-major axis would be of about $0.2\%$ after 100 Myr. 

\subsection{Effect on the resonant pair}

The period ratio of planets b and c in the Kepler 36 system is $P_c/P_b = 1.1733$ and the distance to resonance is $\Delta = P_c/P_b 6/7 -1 = 0.0057 $, using periods listed by \citet{carter12}. If Kepler 36b was once located 0.4\% further away from the star, we find $\Delta=0.0001$, which is much deeper inside the resonance. However, the natural outcome of the torque we have derived here is to move the planet outward, not inward. In principle, inward migration could occur in the case of a retrograde zonal wind. Either, a wind from Kepler 36b did not exert a significant torque on its orbit, or the wind moved the planet only slightly inward rather than outward.

Dissipation in planets from variations in the tidal force from the central star
can slowly cause orbits of planet pairs in mean motion resonance to diverge
\citep{papa11,delisle12,lithwick12,batygin13,lee13}.
Tidal dissipation is a mechanism that could account for the many near resonant
Kepler planet pairs \citep{lithwick12,batygin13,lee13}.
\citet{quillen13} found (following \citealt{lee13}) that it was difficult
to account for the distance of the Kepler 36 system from the 7:6 resonance (assuming it was originally left there)
using tidal forces alone (though the effect of larger planet radii and masses at earlier
times and prior to evaporation was not taken into account).

Some models of tidal evolution predict low eccentricity after escape from resonance \citep[see, e.g.,][with the exception of the scenario by \citealt{delisle12}]{lithwick12}.  However the Kepler 36 system is near some second order resonances and, if there was divergent evolution a resonance crossing could perhaps have lifted the eccentricity of the two planets to near their current values $\sim 0.02$ \citep{carter12}.

While second order resonance crossing (due to drift caused by an evaporative wind) 
presents a possible explanation for Kepler 36's eccentricity, lower index resonances such
as the 3:2 are more distant from second order resonances.  Only if the wind increases
the eccentricity could non-zero eccentricities in lower index resonances be related to
orbital variations associated with an evaporative wind.

\section{Summary and Discussion}

In this paper we have studied the torque exerted by an evaporative wind on planets with properties of super-Earth or hot Neptunes. 

Evaporative winds driven from close-in planet by EUV and X-ray radiation from a  star are unlikely to be isotropic. If the upper atmosphere of a wind undergoing heavy mass loss is super-rotating, then we estimate that there would be a lag on the dawn side between the time the upper atmosphere is illuminated by starlight and when a newly launched wind becomes transonic. This lag results in a torque on the planet.  A planet that loses a significant fraction of its mass could drift by a small amount in semi-major axis. By estimating wind launch timescales we find that only in a narrow regime of planet radius, mass and UV + X-ray flux can the lag angle be significant. Consequently only in rare cases could a planet's orbit (or eccentricity) be affected by the wind's torque. 

Our 1D, time-dependent, hydrodynamical simulations have shown that for a close-in, Neptune-like planet, the anisotropic torque exerted by the wind does not strongly affect the semi-major axis of the planet. For the cases we have considered here, a Neptune-like planet on a short period orbit could undergo a semi-major axis drift of the order of $1\%$. However, this small effect has slightly more significant consequences in the case of Kepler 36b. The small change in semi-major axis that the wind could cause might be important in sculpting the near-resonant configuration of the system. We find that the amount of drifting in semi-major axis caused by an anisotropic wind toque could possibly account for Kepler 36 b,c's distance from the 7:6 resonance.

More generally, this mechanism has important consequences for close-in resonant pairs of planets. If the wind circulates in the prograde direction, the torque arising from the evaporative wind would cause outward migration of the innermost planet. Assuming that the outer planet is less affected (as it receives a weaker flux) then the pair would be pushed deeper inside resonance. Therefore, any mechanism that predicts resonance repulsion should be strong enough to counteract the torque arising from an evaporative wind. 
If the wind circulation was retrograde, then the innermost planet would be pushed inward. We have found that for the case of Kepler 36b, it is enough to bring it to its current location just outside the 7:6 mean motion resonance. Hence, there is strong motivation to do global circulation models for strongly irradiated hot Neptunes. 
We note that for low order resonances, such as the 2:1 resonance, the effect of the torque on $\Delta$ (distance to resonance) might be weak because of the large orbital separation between the two planets (compared to 7:6 resonance, for instance). However, we point out that M type stars have a strong X-ray flux for about 1 Gyr, so the time-integrated effect for planets around such stars will be more important.

The values we have estimated for the lag angle depend on the zonal speed we use (here 1 km/s). If the zonal winds were rotating slower, the lag angle would also be smaller, which would minimize the influence of the torque. In this sense, the lag angle that we derive here can regarded as an upper limit on how an evaporative wind can affect a planet's orbit. Global 3D simulations of evaporative atmospheres could also put more accurate constraints, since our 1+1D simulations can only be considered illustrative of the kind of anisotropic winds that will arise. 3D simulations would not suffer from the uncertainly in the lag angle that arises from our approach.

\vskip 0.2 truein
--------

This work was initiated during the International Summer Institute for
Modeling in Astrophysics (ISIMA) in 2014, hosted at CITA at the University of Toronto. This work was in part supported by NASA grant NNX13AI27G. We also thank Kristen Menou, Linda Strubbe, Pascale Garaud, Jeremy Leconte, Yanqin Wu, Subu Mohanty, Man-Hoi Lee, Eric Lopez and John Papaloizou for helpful discussions and correspondence. JT is supported by STFC through grant ST/L000636/1. JEO acknowledges support by NASA through Hubble Fellowship grant HST-HF2-51346.001-A awarded by the Space Telescope Science Institute, which is operated by the Association of Universities for Research in Astronomy, Inc., for NASA, under contract NAS 5-26555.

{}

\appendix
\section{Orbital evolution}
\label{ap:orbital}

In this section we compute the rate of change of semi-major axis and eccentricity caused by a wind's momentum loss rate.
Following \citet{burns76}, we average the force per unit mass exerted by the wind on the planet over the planet's orbit.

\subsection{Semi-major axis evolution}

A wind that carries momentum at a rate $\dot {\bf P}$ exerts a force per unit mass ${\bf F}_w = \dot {\bf P}/M_p$ on the planet.
Following \citet{burns76} we decompose
the force per unit mass from the wind  into components 
\begin{equation}
{\bf F}_w = T \hat {\boldsymbol \theta} + R \hat{\bf r} + N \hat {\bf z}
\end{equation}
in spherical coordinates defined by the planet's position $r \hat  {\bf r}$ and with 
$\hat {\bf z}$ aligned with the orbital angular momentum
vector.
Here $T, R, N$ have units of acceleration.
The work per unit mass (change in orbital energy per unit mass) is
\begin{equation}
\dot E_o = {\bf v} \cdot {\bf F}_w  = \dot r R + r \dot \theta T. 
\end{equation}
The change in orbital energy directly gives an estimate for the drift rate in semi-major axis
\begin{equation}
\dot a_p =  \frac{2 a_p^2}{GM_*} (\dot r R + r \dot \theta T).
\end{equation}
At low eccentricity, the timescale for drift in semi-major axis is
\begin{equation}
\tau_a \equiv \frac{ a}{\dot a_p} \approx n^{-1} \frac{GM_*}{2 T a^2} \label{eqn:tau_a}
\end{equation}
where $n$ is the planet's mean motion.
Because $\dot a_p$ is non-zero at zero eccentricity,  we ignore the sensitivity of 
the migration rate to eccentricity. 

\subsection{Eccentricity evolution}
\label{ap:ecc}

For the eccentricity evolution
\begin{eqnarray}
n^{-1} \frac{d}{dt} \frac{e^2}{2}
 &=& \frac{2 HE_o}{n} \dot H + \frac{H^2}{n} \dot E_o \nonumber \\ 
&=& \frac{2 HE_o rT }{n} + \frac{H^2 }{n} ( \dot r R + r \dot \theta T) \label{eqn:ee_0}
\end{eqnarray}
where
the angular momentum per unit mass reads
\begin{equation}
{\bf H} = na_p^2 y \hat {\bf z}.
\end{equation}
Here $e$ is the planet's eccentricity and $y \equiv \sqrt{1-e^2}$.
Whereas $\dot a_p$ contains a non-zero eccentricity independent term, eccentricity damping or growth
only takes place if there is eccentricity and both terms in equation \ref{eqn:ee_0} are needed to calculate
this rate.  The torque $\dot {\bf H} = {\bf r} \times {\bf F}$, and the magnitude
$\dot H = r T$ in terms of the tangential force component.
The radius from the star is
\begin{equation}
r= a_p(1-e \cos  E) 
\end{equation}
 with $E$ the eccentric anomaly. The 
 mean anomaly $M$ and the eccentric anomaly are related by Kepler's equation
 \begin{equation}
 M = E - e \sin E. 
 \end{equation}
 The mean motion
 \begin{equation}
 n = \dot E(1-e\cos E).
 \end{equation}
Inserting these two equations into our expression for radius, we get
\begin{equation}
\dot r = a_p e \sin E \dot E = \frac{a_p e \sin E n}{1 - e \cos E}.
\end{equation}
Using 
$H = r^2 \dot \theta$
we find
\begin{equation}
r \dot \theta = \frac{na_p^2 y}{r} = \frac{na_p y}{1 - e \cos E}
\end{equation}

Putting expressions that depend on the eccentric anomaly
into equation (\ref{eqn:ee_0}), we find  for the eccentricity evolution
\begin{multline}
n^{-1} \frac{d}{dt} \frac{e^2}{2} = \frac{a_p^2}{GM_*} \left[
-T(1 - e \cos E) y + \frac{Ty^3}{1 - e \cos E} \right.  \\ \left. + \frac{Re \sin Ey^2}{1-e \cos E}  \right]\\
\end{multline}
This is equivalent to equation (28) by \cite{burns76} but this is written in terms of the eccentric anomaly so it can 
more easily be integrated.

Let us assume that $R$ and $T$ both depend on radius in the same way so that
\begin{equation} 
T \equiv R \sin \theta_w 
\end{equation}
with $\theta_w$ the angle of the integrated outflow momentum.
This means that $T<R$.
\begin{multline}
n^{-1} \frac{d}{dt} \frac{e^2}{2} =  \frac{R a_p^2}{GM_*} \left[
 \sin \theta_w y \left( -(1 - e \cos E)  + \frac{y^2}{1 - e \cos E} \right) \right.  \\  \left. + \frac{e \sin Ey^2}{1-e \cos E}  \right]
 \label{eqn:ee_1}
\end{multline}

Differentiating Kepler's equation
\begin{equation}
dM = n dt = dE (1- e\cos E)
\end{equation}
and we 
average equation (\ref{eqn:ee_1}) over a rotation period
\begin{multline}
\langle n^{-1} \frac{d}{dt} \frac{e^2}{2} \rangle = 
\frac{1}{2\pi} \frac{a_p^2 }{GM_*} \int_0^{2 \pi} {dE}~ R(E) \\
\left[ \sin \theta_w y (2e \cos E - e^2 \cos^2E - e^2) \right.   \\ 
\left.  + e \sin E y^2 \right]  \label{eqn:bare}
\end{multline}
There is no eccentricity damping unless there is some eccentricity.
Hereafter we drop the brackets and discuss the orbit averaged eccentricity evolution.

Assume that with a low order eccentricity expansion we can write
\begin{equation}
R(E) \sim \bar R ( 1 + r_c  \cos E + r_s  \sin E) \label{eqn:RE}
\end{equation}
where $r_c$ and $r_s$ are unit-less but could depend on eccentricity.
If the momentum flux $R(E) \propto r^{-\beta}$ then to first order $R(E) \propto (1- \beta e\cos E)$
and $r_s=0$.
If there is a seasonal lag, then we might get non-zero $r_s$ term.
Integrating equation (\ref{eqn:bare}) we find
\begin{equation}
n^{-1} \frac{d}{dt} \frac{e^2}{2}  \sim \frac{\bar R a_p^2 }{GM_*}e \left( \sin \theta_w \left(  - \frac{3}{2} e +  r_c \right) +  \frac{r_s}{2} 
\right) 
\end{equation}
and we have dropped terms higher than second order in eccentricity because equation (\ref{eqn:RE}) 
is a low order approximation.

If the wind momentum flux depends inversely on radius then $r_c \sim e$ and 
\begin{equation}
n^{-1} \frac{d}{dt} \frac{e^2}{2} \sim -\frac{\bar R a_p^2 }{GM_*} \frac{e^2}{2}  \sin \theta_w   
\end{equation}
and we expect eccentricity damping.
If the wind momentum flux depends inversely on the square of radius then $r_c \sim 2e$ and
\begin{equation}
n^{-1} \frac{d}{dt} \frac{e^2}{2} \sim \frac{\bar R a_p^2 }{GM_*} \frac{e^2}{2}  \sin \theta_w  
\end{equation}
and we expect an increase in eccentricity.
\citet{owen12} predict that X-ray driven winds have mass loss rate proportional to $r^{-2}$ and
EUV driven winds have mass loss rates proportional to $r^{-1}$. This suggests that for
planets within 0.1 AU of a star, the wind could cause the orbital eccentricity 
to increase whereas external to this the eccentricity would be damped.

And in both cases the timescale for eccentricity variation is similar to 
\begin{equation}
\tau_e = \frac {e^2}{ d e^2/dt} \sim \frac{1 } {n \sin\theta_w}  \frac{ GM_* }{\bar R a^2}  
\end{equation}

Our expression for the eccentricity damping (or increase) timescale and eccentricity evolution rate
depend on $\bar R  a_p^2 /(GM_*) = \bar R $ and $\bar R$ is in units of acceleration as it is is force per unit mass.
This implies that 
\begin{equation} 
\bar R \sim \frac{ \dot M v_w} {M_p}
\end{equation}
or 
\begin{equation}
\frac{\bar R a_p^2 }{GM_*} = \frac{v_w}{v_c} \frac{\dot M}{M_pn}
\end{equation}
which depends on the ratio of the wind outflow velocity and orbital speed
and the mass loss rate related to the fraction of mass lost per orbital period.
The torque we previously estimated is $\tau_w = M_p T$, giving
\begin{equation}
\sin \theta_w = \frac{\tau_w }{\dot M v_w}
\end{equation}
Putting these together , we find:
\begin{equation}
\tau_e = \frac {e^2}{ d e^2/dt} = \frac{1 } { \sin\theta_w}  \frac{v_c}{v_w} \frac{M_p}{\dot M}
\end{equation}
and $\tau_e \sim \tau_a$. 
We find that the  ratio commonly used to
describe migration rates,  $K \equiv \tau_a/\tau_e$  \citep{lee02},  
is of order unity for evaporative winds.


\end{document}